# Structural Change Analysis of Active Cryptocurrency Market

Tan, Chia-Yen[1], Koh, You-Beng[2], Ng, Kok-Haur[3]
[1,2,3]Institute of Mathematical Sciences, Faculty of Science, University of Malaya, 50603 Lembah Pantai, Kuala Lumpur, Malaysia.
E-mail: sma170059@siswa.um.edu.my; kohyoubeng@um.edu.my; kokhaur@um.edu.my

Ng, Kooi-Huat [4]
[4]Department of Mathematical and Actuarial Sciences, Lee Kong Chian Faculty of Engineering and Science, Universiti Tunku Abdul Rahman, Malaysia.
Email: khng@utar.edu.my

**Abstract**

Motivated by the large frequent price fluctuation and excessive volatility observed in cryptocurrency market, this study adopts Bai and Perron's (2003) structural change model by incorporating exogenous variables to examine the number and location of change points in daily price, return and volatility proxied by the squared return of Cryptocurrency Index (CRIX), Cryptocurrency Index 30 (CCI30), and the top ten cryptocurrencies ranked according to market capitalisation. Results show that structural changes occur very frequently for the price series, followed by squared return and return series which were consistently observed in between December 2017 to April 2018. In addition, the results also reveal that the two cryptocurrency indices may not be beneficial as an indicator to reflect the whole cryptocurrency market for the entire studied period as these two indices do not display consistent structural change in contrast to the top ten cryptocurrencies that might have significant implications for modelling the cryptocurrency data.

**Keywords:** Structure change; Change point; Cryptocurrency; Index; Return; Volatility.

# 1. Introduction

The world had witnessed an explosive growth in cryptocurrency market in year 2017 mainly for Bitcoin which recorded a considerable price appreciation of approximately 1300% (www.coinmarketcap.com). One of the interesting aspects of cryptocurrency market is that the exceptional large price and volatility fluctuations observed in a short span of time. News had been continuously reporting prices of Bitcoin and altcoins which reached all-time high especially in late 2017 when there was an abrupt surge of interest in the market. This can be noted from some sudden spikes in the total of cryptocurrency market capitalisation and trading volume in terms of U.S. dollar as presented in Figure 1. A large retracement soon occurred that caused the total market capitalisation to reduce by about 50% from the peak recorded in January 2018. Investors are undoubtedly exposed to additional risks in a volatile market, and according to Williams (2014), the volatility of Bitcoin is relatively large as compared to other traditional financial assets such as precious metal, Standard and Poor 500 and the U.S. dollar.

[Figure 1 near here]

Different models have been employed in foregoing studies to estimate the volatility of cryptocurrencies. Katsiampa (2017), for instance, fitted the Bitcoin data to six types of generalised autoregressive conditional heteroskedaticity (GARCH) models, in which the autoregressive- component GARCH was found to be the best model to estimate volatility. There were seven cryptocurrencies considered in Chu et al. (2017): Bitcoin, Dash, Litecoin, Maisafe Coin, Monero, Doge and Ripple which are ranked according to market capitalisation. They had fitted 12 different GARCH models to each cryptocurrency and the results showed that the Glosten-Jagannathan-Runkle (GJR-GARCH) with normal innovation fitted well for Dogecoin, GARCH model with normal innovation for Ripple, and integrated GARCH (IGARCH) model with normal innovation for others. Lahmiri et al. (2018) examined the long-range memory in volatility series of seven Bitcoin markets via fractionally integrated GARCH (FIGARCH). Philip et al. (2018) incorporated the stylised facts displayed by cryptocurrency such as generalised long memory effect, leverage and heavy tails to stochastic volatility model and there were 224 different cryptocurrencies being applied to determine which of these properties exist. Peng et al. (2018) used GARCH, exponential GARCH and GJR-GARCH models with three different innovations: normal, Student t, and skewed Student t distributions to contrast the nine models with Support Vector Regression-GARCH (SVR-GARCH) model. They had come to a conclusion that the estimated mean using the volatility equations from SVR, SVR-GARCH outperforms the other nine GARCH models. One issue arises from these studies is that most of the studies only focus on one single model that can best-fit the entire period of volatility which may not be adequate to capture the abrupt shift observed in cryptocurrency market when the presence of structural break is not taken into consideration. Bauwens et al. (2015) tested the presence of structural breaks in 23 macroeconomic series such as Gross Domestic Product and Consumer Price Index and demonstrated the importance of addressing the presence of structural change in forecasting. They are in the view that, in the presence of structural change, there is no single model can always provide the optimal result in forecasting.

Numerous studies on cryptocurrencies have attempted to segment return series, based on different criteria, into several independent partitions in order to investigate the volatility of data. Urquhart (2016) highlighted that Bitcoin market is inefficient if the sample is in full length of data from 1 August 2010 to 31 July 2016. However, if the data were separated into two sub-samples by imposing a split after the dramatic surge in Bitcoin price in late 2013, it was found that the market inefficiency of Bitcoin was largely attributed to the first sub-sample while the second sub-sample showed an improvement in Bitcoin market efficiency. Bouoiyour and Selmi (2016), however, noticed that despite the remarkable Bitcoin price appreciation over time, its volatility rate remained at a less pronounced rate since January 2015. Hence, they estimated the volatility of Bitcoin over two main periods: before January 2015 and after January 2015 by employing different models to the two sample periods. To verify the break of Bitcoin data around the 2013 price crash, Bouri et al. (2016) tested the structural break in Bitcoin price around the period before modelling the volatility of Bitcoin. They then separated the long data into two segments: before 2013 price crash and after the 2013 market price. Thies and Mólnar (2018) investigated the presence change points in Bitcoin return series using the Bayesian Structural change model. Fourty eight change points on the average return of Bitcoin were detected and the segments were combined that exhibited the same properties into 7 regimes based on their volatility. Bouri et al. (2019) determined structural change of logarithmic Bitcoin price, absolute return and squared return series via Bai and Perron (2003) approach where four and five change points were respectively observed in these series derived from the two different exchange platforms, Bitstamp and Coindesk, in which, a maximum allowable of change in each series is only five. Their findings also encompassed change points of price cash in December 2013. Mensi et al. (2018), on the other hand, revealed the impact of structural change in Bitcoin GARCH estimation and suggested that ignoring the presence of structural changes may lead to the persistency of overestimation in volatility. Nevertheless, there has been thus far relatively little progress on the determination of the number and location of change points in price, return and volatility proxied by squared return for cryptocurrency indices, and altcoins. This is a research gap should be further filled.

Our aims in this paper are threefold. Firstly, as many previous studies tend to focus only on Bitcoin or on a limited number of popular cryptocurrencies, we provide statistical analysis for the top ten cryptocurrencies ranked according to market capitalisation as at 30 April 2018, namely, Bitcoin (BTC), Ethereum (ETH), Ripple (XRP), Bitcoin Cash (BCH), EOS, Cardano (ADA), Litecoin (LTC), Stellar (XLM), IOTA and Tron (TRX). In addition, this study is also extended to the whole cryptocurrency market by advocating the use of cryptocurrency index as an indicator to typify the entire market, in which, the two cryptocurrency indices considered are Cryptocurrency Index (CRIX) and Cryptocurrency Index 30 (CCI30). Price, return and squared return series are the primary inputs in risk management and portfolio selection since the price will reflect the market information while the return and squared return will represent expected gain and volatility in financial modelling and financial applications. As for the factor that influences the dynamics of cryptocurrency data, daily trading volume is often reported to be the significant factor that influences Bitcoin price level (Kristoufek, 2015 and Urquhart, 2018). Secondly, we will apply technique introduced by Bai and Perron (2003) to detect the number and location of change points in price, return and squared return of the series for the cryptocurrencies and indices by incorporating the significant autoregressive terms, and exogenous variables such as daily trading volume in the structural change model. Finally, it is paramount importance to examine the performance of the two indices in representing the cryptocurrency market as a whole. To this end, we would compare and evaluate the consistency of the findings by collating the individual cryptocurrency results with the other two indices.

The rest of the paper is organised as followed. Section 2 introduces the data with the descriptive statistics provided, while Section 3 elaborates on the methodologies of how to identify the significant exogenous variables and of detecting multiple structural changes of the series. Empirical results are discussed in Section 4, and lastly, Section 5 will conclude the paper.

## 2. Data

### 2.1 CRIX and CCI30

Two cryptocurrency indices, CRIX and CCI30 daily closing prices in U.S. dollar, are retrieved from their official websites: http://crix.hu-berlin.de and https://cci30.com. CRIX data spans from 8 August 2014 to 30 April 2018 consisting of 1362 observations were introduced by Trimborn and Härdle (2016). The number of constituents in the index are chosen in steps of five based on a lengthy time-varying selection method that relies on Akaike Information Criterion. CCI30 data on the other hand starts from 9 January 2015 and ends on 30 April 2018 with a total of 1208 observations which was acquired from Rivin and Scevola (2018). The index members are chosen to be the first 30 cryptocurrencies ranked according to market capitalisation. The base number for CRIX and CCI30 are 1000 and 100 respectively. Both indices are rebalanced every month and reconstituted for each quarter.

Return series are obtained by taking the difference of two consecutives log prices and squared return series are computed by taking the square of return which is commonly used as an indicator for variance or risk in financial terminology. Figure 2 below illustrates the price, return and squared return series for CRIX and CCI30. The vertical dotted lines are the estimated change points while the red lines represent the fitted equation estimated using Equation (1) in the presence of change points which is discussed in Section 4. The large fluctuation of cryptocurrency market can be observed by the strikingly sharp increase in price since year 2017 which is soon followed by an abrupt drop in year 2018. This corresponds to the spikes noticed in the return and squared return series indicating by high volatility level for the cryptocurrency market. As seen in Figure 2, the fitted equations (in red colour) are able to capture the prominent spikes observed in the series.

**[Figure 2 near here]**

### 2.2 Top Ten Cryptocurrencies

Daily closing prices in U.S. dollar of the top ten cryptocurrencies were retrieved from Yahoo Finance (https://finance.yahoo.com). These cryptocurrencies constitute 79% of the total market capitalisation with BTC dominating 37.04% of the overall market as at 30 April 2018. This is due to that BTC has been introduced since year 2010 whereas most cryptocurrencies data are introduced in year 2017 onward. Table 1 presents the summary statistics of the indices and daily closing prices for these ten cryptocurrencies. From Table 1, more than 50% of the times the observed closing prices of cryptocurrencies will earn below its expected return. All these indices and cryptocurrencies display positively-skewed and leptokurtic distributions except for ETH, BCH, EOS, ADA, XLM and IOT with platykurtic distributions.

**[Table 1 near here]**

Figure 3 illustrates the price, return and squared return series for each of these cryptocurrencies. The vertical dotted lines represent the estimated change points while the red line is the fitted equation estimated using Equation (1) in the presence of change points. Price series appear to have the most change points compared to return and squared return series. As indicated in Figure 3, the fitted equations (in red colour) are able to capture the changes of the series especially during prominent spikes. The details of change point detection are discussed in Section 4.

**[Figure 3 near here]**

## 3. Methodology

To test the significance of exogenous variables, we commence by fitting the three time series data on the price, return, and squared return under different levels (constants) and daily trading volumes for the respective autoregressive terms of the two cryptocurrency indices and the top 10 cryptocurrencies. For price series, only the level (constant) and trading volume (except for CRIX and CCI30) are considered while for the return and squared return series; level, trading volume and the autoregressive terms are included. Table 2 provides the parameter estimates and the significance of these variables in price, return and squared return series of cryptocurrencies data.

The significant exogenous variables are then incorporated into the multiple structural change model to detect the change points. Liu, Wu and Zidek (1997) who first considered multiple structural change using least-squares method partitioned the data into *m* segments in order to compute the sum of squared residuals for each segment with the idea that the change point estimators used are regarded as the global minimisers of the sum of squared residuals. Bai and Perron (2003) extended their work by applying dynamic programming algorithm to estimate the global minimisers of the sum of squared residuals. The algorithm uses at most least-squares operations of order $O(T^2)$ for any number of *m* change points which appears to be a more efficient way of achieving a minimum global sum of squared residuals with *T* representing the length of data.

Consider a linear model below with *m* changes (*m*+1 segments):

$$y_t = \boldsymbol{x}_t'\boldsymbol{\beta} + \boldsymbol{z}_t'\boldsymbol{\delta}_j + u_t, \quad t = T_{j-1} + 1, T_{j-1} + 2, \dots, T_j,$$

for $j = 1, 2, \dots, m + 1$. In this model, $y_t$ is the observed dependent variable at time *t* with dimension 1 x 1; $\boldsymbol{x}_t'$ is a 1 x $k$ vector of exogenous variables with $k$ number of constant coefficients in vector $\boldsymbol{\beta}$ of dimension $k$ x 1, $\boldsymbol{z}_t'$ is a 1 x $n$ vector of exogenous variables with $n$ number of corresponding coefficients that are subject to change in vector $\boldsymbol{\delta}_j$ of dimension $n$ x 1 and $u_t$ is the disturbance at time $t$ with dimension 1 x 1. The indices $(T_1, T_2, \ldots, T_m)$ are the estimated change points treated as unknown with $T_0 = 0$ and $T_{m+1} = T$. In this study, we concentrate on pure structural change model which allows all the coefficients to be subject to change by letting $\boldsymbol{\beta} = \boldsymbol{0}$ so that the shifts of all exogenous variables, if any, are considered. The above multiple linear regression system can then be expressed in its matrix form as below, with $\overline{\boldsymbol{Z}}$ is a diagonal matrix that partition $\boldsymbol{Z}_j$ at $(T_1, T_2, \ldots, T_m)$:

$$Y = \overline{Z}\delta + U, \qquad (1)$$

or

$$\begin{pmatrix} y_1 \\ y_2 \\ \vdots \\ y_T \end{pmatrix} = \begin{pmatrix} Z_1 & 0 & \cdots & 0 \\ 0 & Z_2 & \cdots & 0 \\ \vdots & \vdots & \ddots & \vdots \\ 0 & 0 & \cdots & Z_{m+1} \end{pmatrix} \begin{pmatrix} \delta_1 \\ \delta_2 \\ \vdots \\ \delta_{m+1} \end{pmatrix} + \begin{pmatrix} u_1 \\ u_2 \\ \vdots \\ u_T \end{pmatrix} \quad \text{where } Z_j = \begin{pmatrix} z_{T_{j-1}+1}' \\ z_{T_{j-1}+2}' \\ \vdots \\ z_{T_j}' \end{pmatrix}$$

where the dimension for $Y$ is $T$ x 1, for $\overline{Z}$ is $T$ x $n(m+1)$, for $\delta$ is $n(m+1)$ x 1 and for $U$ is $T$ x 1.

For each $m$-partition $(T_1, T_2, \ldots, T_m)$ denoted by $\{T_j\}$, the estimates of $\boldsymbol{\delta}_j$ are evaluated by minimising the sum of squared residuals. Substituting the values into the objective function and denoting the resulting sum of squared residuals as $S_T(T_1, T_2, \ldots, T_m)$, the estimated change points $(\hat{T}_1, \hat{T}_2, \ldots, \hat{T}_m)$ are determined by $(\hat{T}_1, \hat{T}_2, \ldots, \hat{T}_m) = argmin_{T_1, \ldots, T_m} S_T(T_1, T_2, \ldots, T_m)$, where the minimisation is taken over all partitions $(T_1, T_2, \ldots, T_m)$. Consequently, the change point estimators are the global minimisers of the objective function. The global sum of squared residuals for any $m$-partition $(T_1, T_2, \ldots, T_m)$ and for any value of $m$, must necessarily be a particular linear combination of these $T(T+1)/2$ sums of squared residuals. The estimates of the change points, $m$-partitions $(\hat{T}_1, \hat{T}_2, \ldots, \hat{T}_m)$, will correspond to this linear combination with a minimal value. The dynamic programming algorithm is regarded as a more efficient approach to contrast all possible combinations (corresponding to different $m$-partitions) in order to achieve a minimum global sum of squared residuals. The number of changes are controlled by the trimming error, $h$, where $h$ is the ratio of number of days in a segment over the total number of days (count). In this study, we let $h$ to be the smallest possible value so that we would be able to detect the presence of change points even if the changes occur in a short span of time without limiting the number of changes. For estimation purposes, the number of days in a segment must be greater than the number of regressors in the model. It is worth noting that, Bouri et al. (2019) would only allow a maximum of five changes in the detection of structural change process. The above algorithms are implemented using the R package **strucchange** with function **breakpoints**.

Table 2 presents the parameter estimates of trading volume and autoregressive terms on price, return and squared return series for different types of the cryptocurrencies. Our findings reveal that the trading volume indeed has a significant impact on price series for all cryptocurrencies except for ADA. With regards to the return series, trading volume is a significant exogenous variable for XRP, BCH, EOS, ADA, LTC, XLM, and for squared return series, the trading volume is a significant exogenous variable for all cryptocurrencies except for BTC, XRP and TRX. It is noticed that trading volume has impact merely on BTC price series but not for the return and squared series. This result can further be substantiated by Balcilar et al. (2017) who had showed that the trading volume can be used to predict Bitcoin return only when the market is fluctuating around the median but not viable when the market is either bearish or bullish. Moreover, they also highlighted that the volatility of Bitcoin is not appropriate for prediction purposes through trading volume.

**[Table 2 near here]**

### 4. Empirical Analysis

*4.1 Price series*

Price series is greatly affected by the force of supply and demand and other external events. It reacts sensitively to news and information in the market. When demand is greater than supply, the price will ascend and the market is bullish or vice versa. It is noticed that there were few consistent and apparent change points occurred in the individual cryptocurrencies over time. Figure 4 depicts the monthly segmentation of price series, in which a change point is represented by an alternate change in colour of the horizontal bar.

**[Figure 4 near here]**

Results reveal that change points in price series are detected specifically at the turning of year. Observing the lengthier data such as CRIX, CCI30, BTC and LTC, the changes were detected almost at every ending or beginning of the year commencing from year 2013 when cryptocurrency market begins to gain its popularity. We hence postulate that the cryptocurrency market is subject to "year-end" effect as there are cyclical changes in price at the turnings of years. Our results also provide evidence and justification in terms of the location of change point as indicated in the study conducted by Bouri et al. (2016) which happened in BTC price crash of December 2013. It is also confirmed that the cryptocurrency market indeed underwent unexpected high price fluctuations in 2017 supported by the presence of change points in approximately each month of 2017 for all data. To be more concrete, Appendix A.1 presents the detailed number of the detected change points in the respective months throughout the whole sampled period for the price series. Appendix A.1 will be provided upon request.

*4.2 Return series*

Return is an important variable in finance as it measures the profit of an investment. High return will usually be accompanied by high risk, hence, in the event of high volatility, high return is also anticipated. Figure 5 illustrates the monthly segmentation of return series where a change point is represented by an alternate change in colour of the horizontal bar.

**[Figure 5 near here]**

From Figure 5, there was no change point in BTC return series after the second quarter of year 2014. We also notice that the ETH return series was detected with only one change point located in September 2015 and none thereafter. On the other hand, no change point was detected for TRX in the return series.

Among all the change points identified, CRIX and CCI30 consistently indicated the change points in January 2015 and September 2017. These estimated change points may be attributed to some unforeseen events affecting some groups of cryptocurrencies which were not observable. Moreover, this may also due to several members of the indices that were reconstituted at every quarter and the ten individual cryptocurrencies might not be the components of the indices at those particular periods of time. Besides, CRIX return series was also detected with one change point in December 2017 while the change point of CCI30 return series appeared in late January 2018 during the abrupt market price depreciation. Both change points were to be expected since the cryptocurrency market was in the process of the change of its trend towards the beginning of years 2017 and 2018 as most of the altcoins also displayed the analogous patterns in this period. Appendix A.2 provides the number of detected change points in the corresponding months throughout the whole sampled periods for the return series.

### 4.3 Squared return series

Squared return is commonly used to assess the uncertainty of the financial market. In this study, we also approximate the volatility by using squared return estimates and Figure 6 illustrates the monthly segmentation of squared return series in which a change point is represented by an alternate change in colour of the horizontal bar.

[Figure 6 near here]

As can be seen in the Figure 6, there are more change points detected in squared return series as compared to return series especially for ETH where there is only one change point detected in return series, and yet, there were 23 change points in total for squared return series throughout the period (see Appendix A.3). BTC appears to experience more change points in squared return series in 2011 and tends to become lesser from year 2012 to year 2017 with that of no change point detected in year 2016. Our finding is in line with the selection of change point at beginning of year 2015 by Bouoiyour and Selmi (2016) that further confirms volatility change of BTC at that period. Meanwhile five change points were detected for ETH squared return series whereas most altcoins showed the presence of change points in July 2017, the last quarter of year 2017 and in January 2018. There were also change points detected in squared return series for both indices which showed consistent signs of change in January 2015, July 2017, and in the turning point of years 2017/2018. Appendix A.3 exhibits the number of detected change points in the respective months throughout the whole sampled period for the squared return series.

### 4.4 Discussion

As indicated in results, there were large numbers of change points detected in price, return and squared return series over the period and it is observed that the detected change points are not totally consistent among the three series with greater number of change points detected in price series, followed by squared return series and the least in return series. Table 3 summarises the total number of change points detected according to the types of series and cryptocurrencies.

[Table 3 near here]

It is also noticed that the number of change points detected in CRIX and CCI30 are not totally consistent with the number of change points identified in the top ten individual cryptocurrencies. Although CRIX and CCI30 comprise a moderate number of selected cryptocurrencies, the constituents are re-evaluated and re-selected each quarter, and hence, the entry of the indices are not the same throughout the period. ElBahrawy et al. (2017) measured the average rank occupation time of cryptocurrencies based on its market capitalisation, and they seemed to warrant two conclusions that the turnover rate of cryptocurrencies are high and the cryptocurrencies keep changing their position in ranks according to market capitalisation. Consequently, the difference of change points detected between the two indices and the individual cryptocurrencies in this study may be due to the high mobility of cryptocurrencies, in which, the 10 individual cryptocurrencies might not be the constituents of CRIX and CCI30 at every point of time throughout the sampled period. Because of its fast-changing position nature, we are in the view that cryptocurrency index may act as a benchmark for the market at a certain point of time only and may not be optimal to be used as an indicator to represent the entire market for a full length of the sampled period.

The primary issues of the abrupt change in cryptocurrency market in the turnings of year 2017/2018 discussed earlier might be due to the large correction of sharp price appreciation, stricter regulations and government involvement, rumours and negative news, and other technological difficulties. These are the influential factors that may contribute to the instability of the market that require the investors and financial practitioners to be more cautious in cryptocurrency market.

*5. Conclusion*

Our empirical findings signify the existence of change points in price, return and squared return series of cryptocurrency data and most of them are not consistent among the three series with most change points detected in the price series and the least change points detected in the return series. Moreover, the locations for data segmentation carried out in some previous related literatures seem to be in line with our estimated change points. Hence the results may indicate to financial practitioners or researchers to realize about the possible instability of parameters that may exist in all aspects of cryptocurrency analysis and modelling process due to the frequent existence of change points affected by underlying internal or external factors.

With the growing interests in cryptocurrency market, one aspect of our concern is whether total trading volume is a crucial factor that will influence the price, return and squared return series. In this case, we are in the view that the volume does have significant impact on the three series, however, different cryptocurrencies show different levels of significance. As to our study, only BTC price series tends to be significant but not for the return and squared return series.

Additionally, we can see that even though the two cryptocurrency indices, CRIX and CCI30, both consist of a moderate number of cryptocurrencies based on its respective selection methods, the indices do not well-capture the properties of the whole market for the entire sampled period. The locations of change point detected are not consistently close to the change points detected in the top ten cryptocurrencies which may merely imply that the indices are not optimal to represent the cryptocurrency market for those periods partly due to its fast-changing nature.

On the basis of the results, integrating the change point method into financial modelling appears to be beneficial in prediction process, specifically when one who wants to model the return or volatility in cryptocurrencies. These findings lend support to the wide applications in financial modelling to accommodate in response to the respective change points Despite the encouraging results of this study as to positive effect of points detected, perhaps an important area for future research in the years to come will be in the refinement of approaches to the analysis of threshold model, structural change model or regime switching model.

**Table 1. Summary statistics of the top ten cryptocurrencies prices.**

| Name | Data Starts from | Count | Min | 1st Q | Median | Mean | 3rd Q | Max | Kurtosis | Skewness |
|---|---|---|---|---|---|---|---|---|---|---|
| **CRIX** | 8/8/2014 | 1362 | 342.10 | 601.60 | 1082.90 | 6193.30 | 6510.50 | 62895.30 | 5.9696 | 2.5057 |
| **CCI30** | 9/1/2015 | 1209 | 57.72 | 96.51 | 266.16 | 2070.19 | 2941.65 | 20796.64 | 5.5487 | 2.3828 |
| **BTC** | 1/8/2010 | 2828 | 0.06 | 11.00 | 263.07 | 1,139.02 | 636.40 | 19,345.49 | 13.6777 | 3.6072 |
| **ETH** | 13/8/2015 | 992 | 0.42 | 9.22 | 12.95 | 183.32 | 297.53 | 1,385.02 | 2.7167 | 1.8132 |
| **XRP** | 28/1/2015 | 1190 | 0.00 | 0.01 | 0.01 | 0.17 | 0.20 | 2.78 | 14.8336 | 3.5119 |
| **BCH** | 7/8/2017 | 267 | 220.30 | 546.70 | 994.70 | 1091.30 | 1407.50 | 3,715.90 | 0.6625 | 1.0250 |
| **EOS** | 5/7/2017 | 300 | 0.49 | 1.30 | 3.41 | 5.07 | 8.30 | 21.41 | 0.1139 | 0.8923 |
| **ADA** | 7/10/2017 | 206 | 0.02 | 0.03 | 0.21 | 0.27 | 0.38 | 1.13 | 1.1132 | 1.1920 |
| **LTC** | 31/10/2013 | 1644 | 1.12 | 3.48 | 4.10 | 28.29 | 22.56 | 357.51 | 8.8048 | 2.9623 |
| **XLM** | 24/1/2017 | 463 | 0.00 | 0.01 | 0.03 | 0.13 | 0.23 | 0.88 | 1.3644 | 1.4965 |
| **IOT** | 19/6/2017 | 316 | 0.16 | 0.50 | 1.00 | 1.44 | 1.96 | 5.32 | 0.4743 | 1.1870 |
| **TRX** | 6/10/2017 | 207 | 0.00 | 0.00 | 0.04 | 0.04 | 0.05 | 0.30 | 11.2817 | 2.6447 |

**Table 2. Parameter estimates on various series and cryptocurrencies with values in parentheses are standard errors of parameter estimates.**

| | Price Series | | | | | | | | | | | |
|---|---|---|---|---|---|---|---|---|---|---|---|---|
| | CRIX | CCI30 | BTC | ETH | XRP | BCH | EOS | ADA | LTC | XLM | IOTA | TRX |
| $h$, Size | **0.0015** | **0.0018** | **0.0015** | **0.0035** | **0.003** | **0.012** | **0.012** | **0.010** | **0.002** | **0.008** | **0.010** | **0.020** |
| Constant | 6193.3* (307.9) | 2070.2* (107.1) | 398.3* (2.3610) | 6.6930* (6.0350) | 0.0959* (0.0071) | 83.85* (4.150) | 3.100* (0.2096) | 0.2807* (0.0018) | 1.8450* (1.0910) | 0.0624* (0.0075) | 0.9524* (0.0537) | 0.0182* (0.0022) |
| Trading Volume | ----- | ----- | 5.79e-06* (5.40e-08) | 9.81e-07* (2.35e-08) | 2.35e-09* (5.58e-11) | 2.29e-06* (1.99e-07) | 4.37e-08* (2.39e-09) | 5.87e-11 (4.29e-10) | 3.12e-07* (9.17e-09) | 1.63e-08* (9.60e-10) | 1.30e-08* (7.29e-10) | 1.07e-08* (6.07e-10) |
| | Return Series | | | | | | | | | | | |
| | CRIX | CCI30 | BTC | ETH | XRP | BCH | EOS | ADA | LTC | XLM | IOTA | TRX |
| $h$, Size | **0.0025** | **0.0035** | **0.0022** | **0.005** | **0.005** | **0.017** | **0.010** | **0.020** | **0.004** | **0.008** | **0.0065** | **0.025** |
| Constant | -0.0022* (0.0011) | 0.0031* (0.0012) | 0.0040* (0.0013) | 0.0072* (0.0027) | 0.0016 (0.0033) | -1.27e-02 (7.30e-03) | -0.0050 (0.0069) | -7.24e-04 (0.0093) | 4.96e-05 (0.0020) | -0.0026 (0.0063) | 0.0046 (0.0066) | 0.0037 (0.0134) |
| Trading Volume | ----- | ----- | -4.40e-12 (3.01e-12) | -1.08e-11 (1.04e-11) | 8.50e-11* (2.64e-11) | 1.62e-10* (3.85e-11) | 2.49e-10* (8.52e-11) | 1.12e-09* (2.57e-10) | 7.63e-11* (1.74e-11) | 3.46e-09* (8.88e-10) | -2.55e-11 (9.54e-11) | 4.69e-09 (3.92e-09) |
| Lag1 | -0.0141 (0.0270) | -0.0065 (0.0286) | 0.0462* (0.0188) | -0.0637* (0.0315) | -0.3281* (0.0291) | 4.61e-02 (0.0617) | -0.0410 (0.0589) | -0.1376 (0.0732) | -0.1280* (0.0247) | 0.0026 (0.0048) | 5.91e-02 (5.73e-02) | -0.0022 (0.0707) |
| Lag2 | -0.0160 (0.0270) | 0.0200 (0.0286) | -0.1693* (0.0188) | 0.0060 (0.0313) | 0.0629* (0.0304) | -0.1400* (0.0620) | 0.0106 (0.0581) | 0.1669* (0.0680) | -0.0542* (0.0247) | -0.0064 (0.0467) | -0.0354 (0.0574) | 0.1772* (0.0709) |
| Lag3 | 0.0400 (0.0270) | 0.0640* (0.0287) | 0.0228 (0.0190) | 0.0707* (0.0313) | -0.0211 (0.0304) | -0.0084 (0.0602) | 0.0398 (0.0572) | 0.0325 (0.0713) | 0.0160 (0.0247) | -0.0027 (0.0463) | 0.1107 (0.0578) | 0.2051* (0.0709) |
| Lag4 | 0.0095 (0.0271) | -0.0069 (0.0287) | 0.0107* (0.0190) | -0.0151 (0.0312) | 0.0191 (0.0304) | -0.1122 (0.0586) | -0.0576 (0.0494) | -0.0159 (0.0693) | 0.0407 (0.0248) | -0.0778 (0.0456) | 0.0057 (0.0572) | -0.1390* (0.0702) |
| Lag5 | 0.0127 (0.0272) | 0.0190 (0.0286) | 0.0835* (0.0190) | 0.0013 (0.0312) | -0.0435 (0.0304) | 0.0201 (0.0575) | -0.0238 (0.0496) | -0.0929 (0.0681) | 0.0032 (0.0247) | 0.0788 (0.0463) | 0.0188 (0.0555) | 0.0179 (0.0697) |
| Lag6 | 0.1055* (0.0272) | 0.1205* (0.0286) | -0.0093 (0.0190) | -0.0009 (0.0293) | 0.0492 (0.0289) | 0.0154 (0.0571) | 0.0602 (0.0493) | -0.0645 (0.0682) | 0.0870* (0.0246) | 0.0152 (0.0459) | 0.0992 (0.0553) | -0.0370 (0.0695) |
| | Squared Return Series | | | | | | | | | | | |
| | CRIX | CCI30 | BTC | ETH | XRP | BCH | EOS | ADA | LTC | XLM | IOTA | TRX |
| $h$, Size | **0.003** | **0.005** | **0.002** | **0.008** | **0.008** | **0.025** | **0.012** | **0.030** | **0.004** | **0.010** | **0.010** | **0.015** |
| Constant | 0.0008* (0.0001) | 0.0008* (0.0015) | 0.0021* (0.0008) | 0.0029* (0.0005) | 0.0057* (0.0018) | 0.0022* (0.0017) | 0.0650* (0.0016) | 0.0112* (0.0045) | 0.0032* (0.0009) | 0.0044* (0.0021) | 0.0055* (0.0016) | 0.0112 (0.0062) |
| Trading Volume | ----- | ----- | 6.19e-13 (1.96e-12) | 3.50e-12* (1.75e-12) | 1.76e-11 (1.38e-11) | 6.61e-11* (8.31e-12) | 9.24e-11* (1.64e-11) | 9.64e-10* (1.27e-10) | 1.55e-11* (7.33e-12) | 1.26e-09* (2.69e-10) | 9.08e-11* (1.98e-11) | 2.70e-09 (1.55e-09) |
| Lag1 | 0.2657* (0.0272) | 0.1821* (0.0284) | 0.0279 (0.0186) | 0.2042* (0.0317) | 0.4496* (0.0291) | 0.1804* (0.0582) | -0.0648 (0.0570) | -0.1591* (0.0733) | 0.5157* (0.0248) | 0.3408* (0.0469) | 0.1065 (0.0565) | 0.0396 (0.0709) |
| Lag2 | 0.0154 (0.0281) | 0.0373 (0.0287) | 0.1406* (0.0186) | 0.0379 (0.0309) | -0.0275 (0.0318) | -0.2546* (0.0594) | -0.0562 (0.0559) | 0.1658* (0.0636) | -0.2092* (0.0278) | 0.0596 (0.0489) | -0.0262 (0.0568) | 0.2049* (0.0710) |
| Lag3 | 0.0001 (0.0281) | 0.0613* (0.0286) | 0.0087 (0.0183) | 0.0747* (0.0308) | 0.1375* (0.0318) | 0.0258 (0.0580) | 0.0003 (0.0056) | -0.1959* (0.0671) | 0.0911* (0.0283) | 0.0064 (0.0488) | 0.0837 (0.0565) | 0.1204 (0.0719) |
| Lag4 | 0.1012* (0.0279) | 0.1124* (0.0286) | 0.2144* (0.0183) | 0.0642* (0.0307) | -0.0519 (0.0318) | -0.0413 (0.0515) | 0.0117 (0.0189) | -0.0191 (0.0647) | 0.0085 (0.0283) | -0.0620 (0.0487) | 0.0004 (0.0560) | 0.0789 (0.0714) |
| Lag5 | 0.0489 (0.0281) | 0.0974* (0.0287) | -0.0070 (0.0186) | 0.0519 (0.0300) | 0.0827* (0.0318) | -0.0046 (0.0512) | 0.0244 (0.0180) | -0.0699 (0.0633) | 0.0019 (0.0278) | 0.0877 (0.0488) | -0.0098 (0.0518) | -0.0766 (0.0700) |
| Lag6 | 0.0118 (0.0282) | 0.0375 (0.0283) | 0.1664* (0.0186) | -0.0305* (0.0143) | -0.0238 (0.0291) | 0.0950* (0.0477) | 0.0116 (0.0178) | -0.1064 (0.0638) | 0.0223 (0.0247) | -0.0779 (0.0455) | 0.0332 (0.0511) | 0.0354 (0.0700) |

----- no data available
\* 5% significance level.

Table 3. Summary of total number of change points detected in price, return and squared return series.

| Cryptocurrency | CRIX | CCI30 | BTC | ETH | XRP | BCH | EOS | ADA | LTC | XLM | IOTA | TRX |
|---|---|---|---|---|---|---|---|---|---|---|---|---|
| Price | 72 | 68 | 72 | 62 | 45 | 27 | 22 | 26 | 66 | 31 | 27 | 10 |
| Return | 7 | 5 | 22 | 1 | 13 | 1 | 1 | 3 | 14 | 4 | 2 | 0 |
| Squared Return | 31 | 19 | 56 | 23 | 8 | 14 | 24 | 9 | 14 | 20 | 16 | 12 |

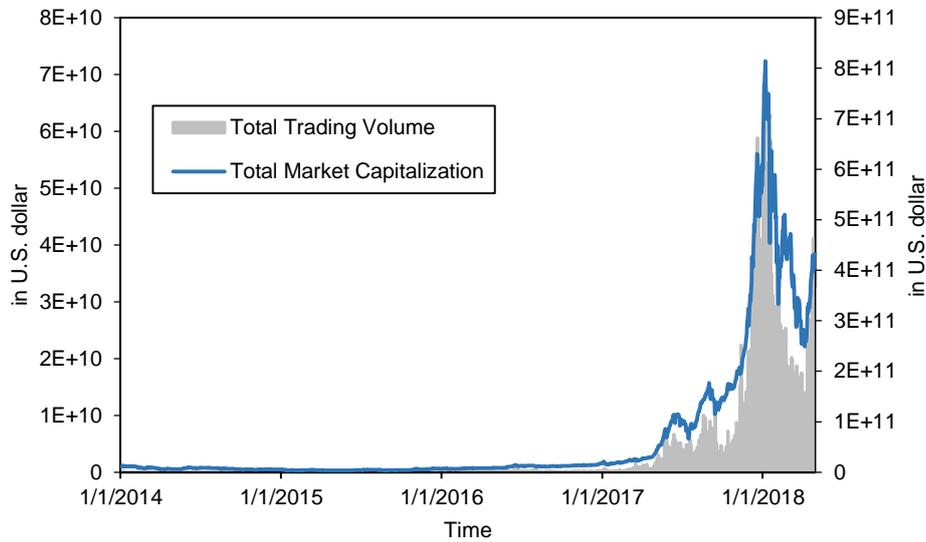

**Figure 1. Total market capitalisation and total trading volume for cryptocurrency market from Jan 2014 to April 2018.**
Source: Data obtained from www.coinmarketcap.com

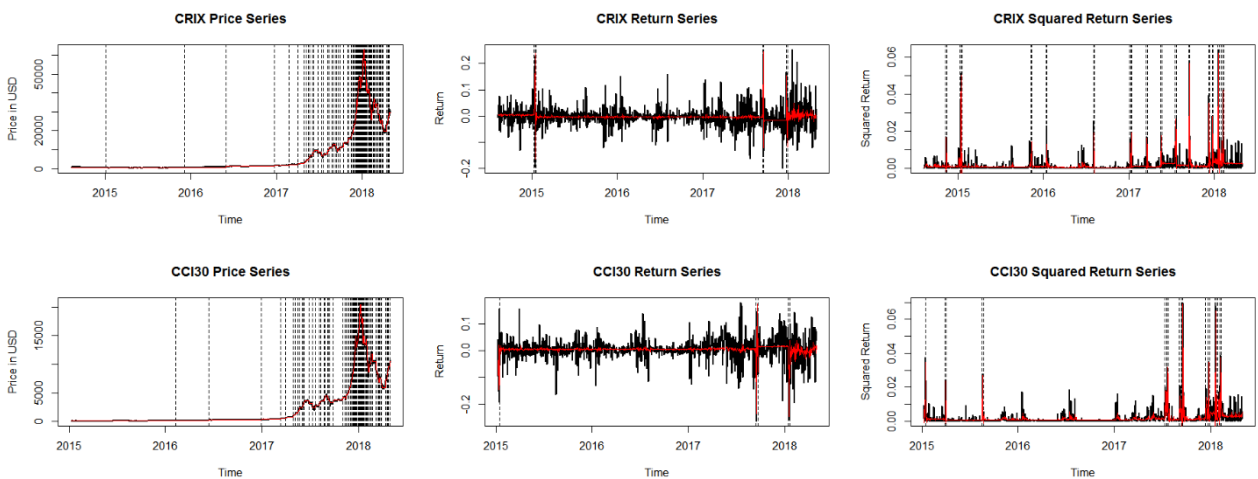

**Figure 2. Price, return and squared return series for CRIX and CCI30.**
Source: Data obtained from http://crix.hu-berlin.de and https://cci30.com

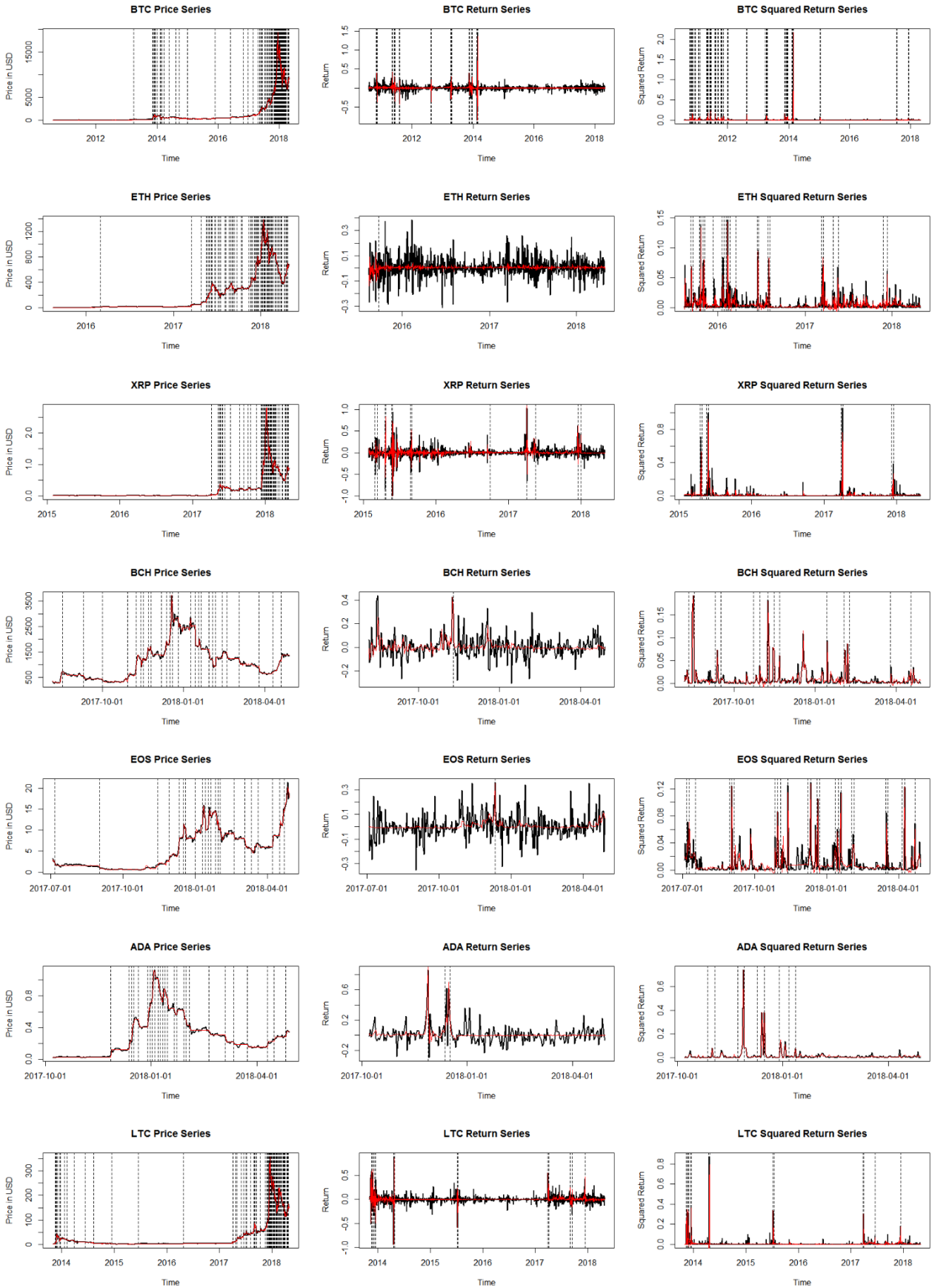

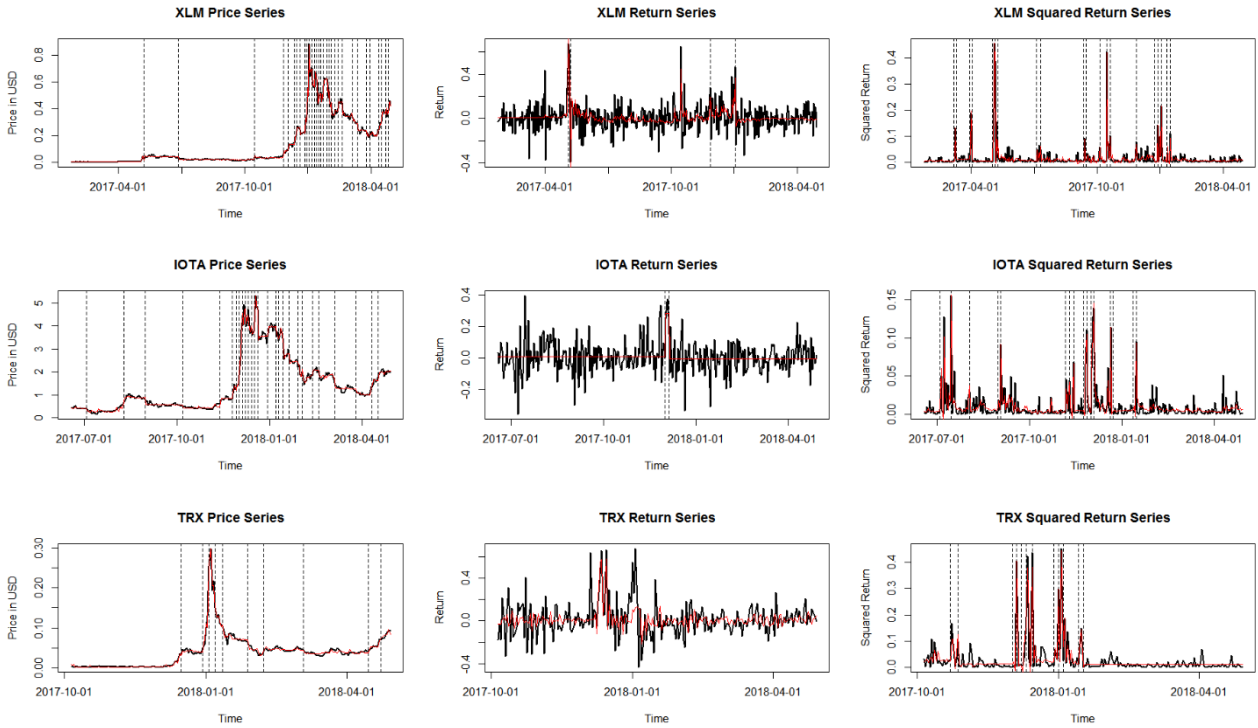

**Figure 3. Price, return and squared return series for the top ten cryptocurrencies.**
Source: Data from https://finance.yahoo.com

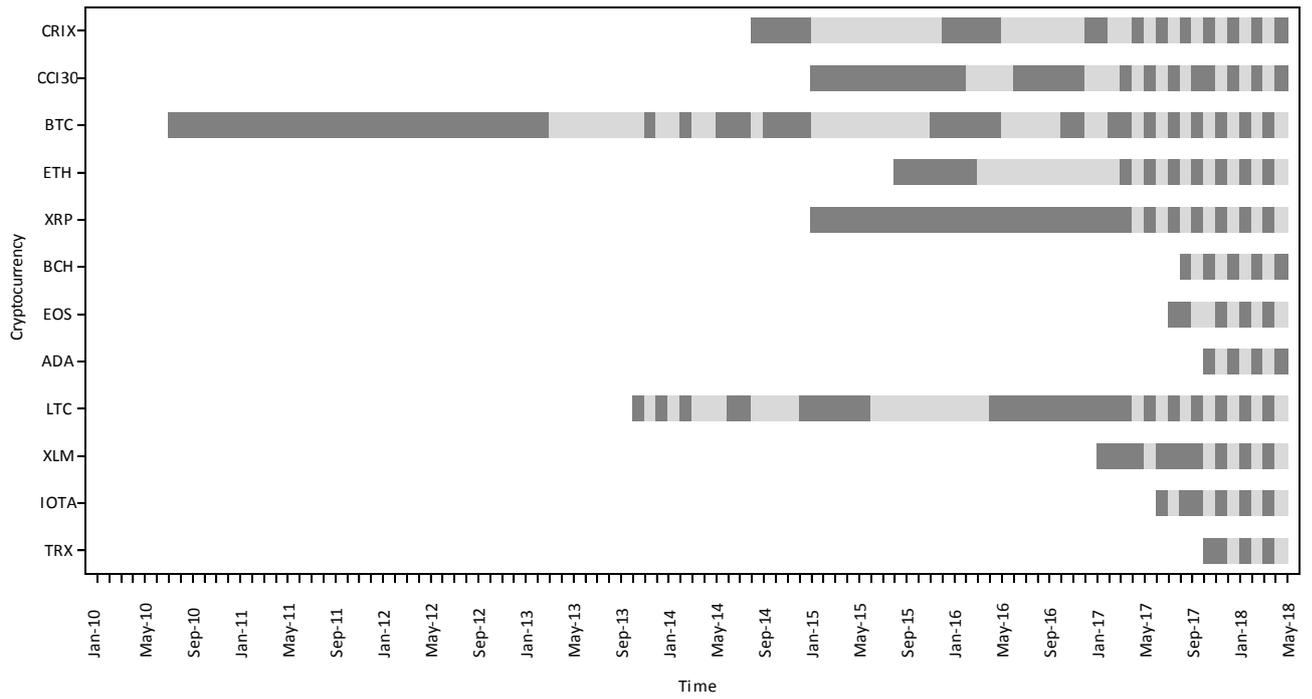

**Figure 4. Locations of change point detected in price series.**

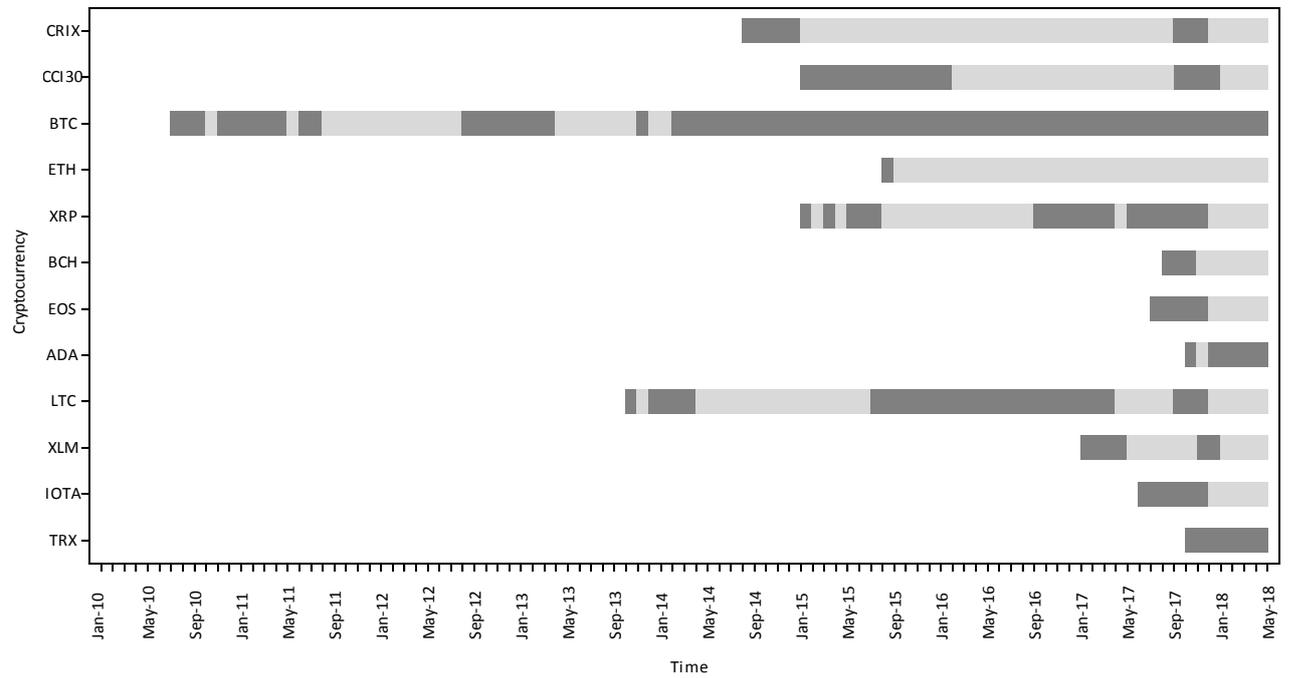

**Figure 5. Locations of change point detected in return series.**

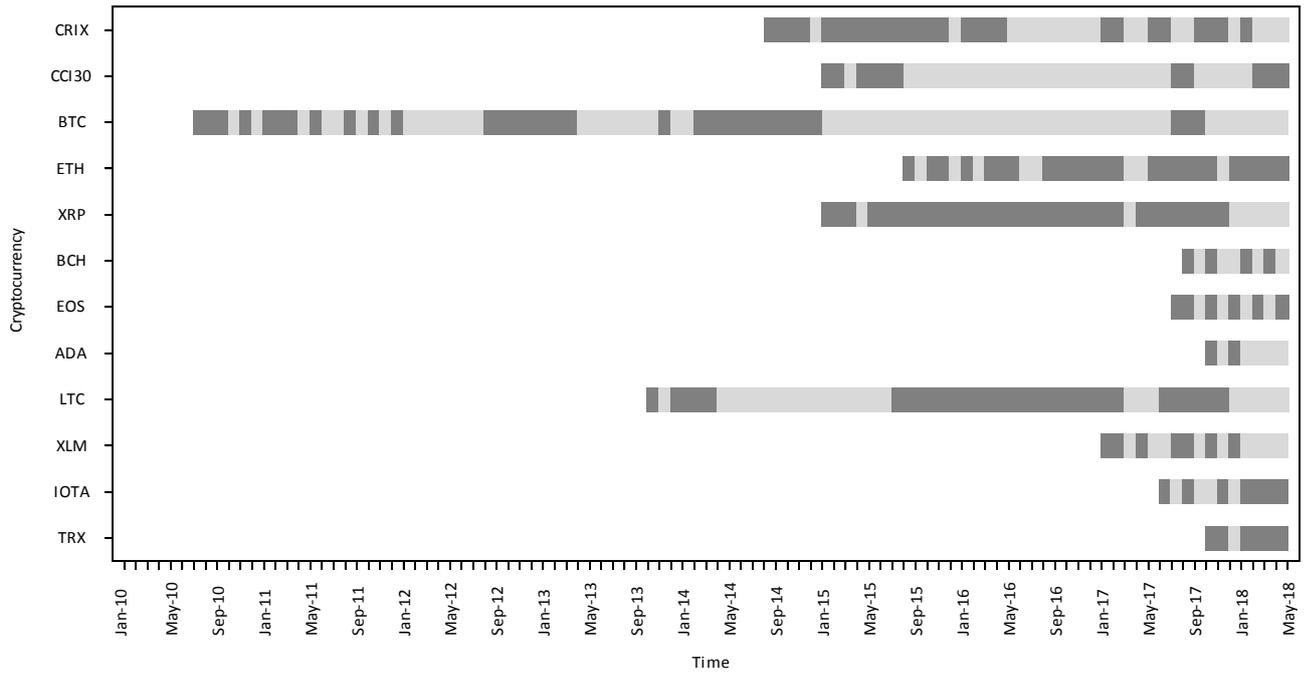

**Figure 6.** Locations of change point detected in squared return series.